# Deconstructing Odorant Identity via Primacy in Dual Networks

Daniel Kepple[1], Hamza Giaffar[1], Dmitry Rinberg[2], and Alexei Koulakov[1]

[1] *Cold Spring Harbor Laboratory, Cold Spring Harbor, NY*
[2] *NYU Neuroscience Institute, NYU Langone Medical center, New York, NY*

## Abstract

In the olfactory system, odor percepts retain their identity despite substantial variations in concentration, timing, and background. We propose a novel strategy for encoding intensity-invariant stimuli identity that is based on representing relative rather than absolute values of the stimulus features. Because, in this scheme, stimulus identity depends on relative amplitudes of stimulus features, identity becomes invariant with respect to variations in intensity and monotonous non-linearities of neuronal responses. In the olfactory system, stimulus identity can be represented by the identities of the p strongest responding odorant receptor types out of a species dependent complement. We show that this information is sufficient to recover sparse stimuli (odorants) via elastic net loss minimization. Such a minimization has to be performed under constraints imposed by the relationships between stimulus features. We map this problem onto the dual problem of minimizing a functional of Lagrange multipliers. The dual problem, in turn, can be solved by a neural network whose Lyapunov function represents the dual Lagrangian. We thus propose that networks in the piriform cortex compute odorant identity and implement dual computations with the sparse activities of individual neurons representing the Lagrange multipliers.

## 1 Introduction

Sensory systems face the problem of computing stimulus identity invariant to several features. The olfactory system, for example, has to compute stimulus identity despite substantial variation in the absolute concentrations of molecules present in the stimulus. This computation is necessary to enable navigation in chemical gradients or within variable odorant plumes. How can the olfactory system robustly represent odorant identity despite variable stimulus intensity? The first step of olfactory processing involves odorants binding to and activating a set of molecular sensors known as olfactory receptors. Olfactory receptors are proteins expressed by olfactory sensory neurons (OSNs) located in the olfactory epithelium. Most mammalian olfactory systems contain ~1000 types olfactory receptors, while humans rely on the responses of only 350 [1; 2; 3]. Importantly, every OSN expresses only a single type of olfactory receptor chosen randomly out of the large ensemble. Odorant identity is therefore represented in the patterns of activation of olfactory receptor proteins and, by extension, OSNs. How the olfactory system represents odorant identity is not clearly understood. Here we examine the hypothesis that stimulus identity is inferred on the basis of the relative amplitudes of responses of OSNs. In particular, we propose that odorant identity can be defined by specifying the p strongest responding. We call this type of representation the primacy model. According to this primacy coding scheme, in which the olfactory system uses relative rather than absolute receptor responses, odorant representations become independent of absolute odorant concentration, leading to a concentration-invariant representation of odor identity. We formulate the identity decoding scheme using a dual Lagrange-Karush-Kuhn-Tucker problem. We show that this dual problem can be solved by a neural network that we call a dual network. Dual networks that solve the primacy problem share many features with real olfactory networks. We therefore derive gross structure of olfactory circuits from first principles, based on the primacy model.

## 2 Dual Networks

### 2.1 Representing odorants by sparse vectors.

Ethologically important odorant stimuli are mixtures of monomolecular components. Such stimuli can be represented by a vector of concentrations $\vec{x}$. Each component of this vector $x_j$ is equal to the concentration of an individual monomolecular component numbered by the index $j$. The number of potential monomolecular components, which is equal to the dimensionality of vector $\vec{x}$, will be denoted here by $M$. This number has been estimated to be on the order of several million, $M \sim 10^6$, based on the count of potentially volatile molecules lighter than 300 Daltons in the popular database PubChem [4]. However, odorant mixtures cannot contain all of these molecules at the same time and are necessarily represented by sparse vectors $\vec{x}$. For the purposes of the olfactory system, the sparseness of the concentration vector is further increased by the inability of the system to detect or recognize individual components. Indeed, psychophysical studies suggest that human observers can detect roughly 12 monomolecular components of the vector $\vec{x}$ [5]. Therefore, we suggest that realistic odorants can be defined by concentration vectors $\vec{x}$ of high dimensionality ($M \sim 10^6$) and very few non-zero elements ($K \sim 10$).

Odorant mixtures are further represented by the responses of olfactory receptor neurons $\vec{r}$. These responses can be approximated by linear nonlinear functions of vector $\vec{x}$. Indeed, in the simplest model of receptor with a single binding site and no cooperativity, the law of mass action yields

$$y_i = \sum_j A_{ij} x_j, \quad r_i = F(y_i) \tag{1}$$

Here index $i = 1...N$ enumerates the olfactory receptor types. The total number of olfactory receptors $N$ varies between 350 in humans and 1100 in rodents. Matrix $A_{ij}$ contains affinities of molecules of type $j$ to the receptors of type $i$. $F(y) = y/(1+y)$ is the nonlinear function describing the activation of a receptor. The problem solved by the olfactory system can be formulated as follows: find the odorant stimulus $x_j$ given the set of responses of olfactory sensory neurons $r_i$. Because the responses $r_i$ are related to their inputs $y_i$ via a simple monotonic function $F$, we can assume that networks analyzing responses of receptor neurons have the linear component of response $y_i$ available to them. Overall, the problem of olfactory decoding is this: find the sparse vector $x_j$ given vector $y_i$ containing olfactory receptor responses.

### 2.2 Sparse olfactory stimulus recovery.

The problem formulated above (find $\vec{x}$ given $\vec{y}$) can be reduced to solving the system of linear equations (1). This problem is not entirely trivial, because the number of unknowns (components of $\vec{x}$, $M \sim 10^6$) is substantially larger than the number of equations (components of $\vec{y}$, $N \sim 10^3$). Some relief arrives from the fact that the vector of unknowns is sparse. A sparse vector can be recovered from a system of linear equations using arguments from compressed sensing [6; 7; 8]. The vector $\vec{x}$ can be found exactly, despite containing more components than equations. To determine $\vec{x}$ given $\vec{y}$, one can use the method of sparse signal recovery via minimization of the $l_1$ norm [6; 7; 8]:

$$\vec{x} = \underset{\substack{\vec{y}=A\vec{x} \\ x_j \geq 0}}{\arg\min} \sum_j x_j \qquad (2)$$

To be able to reconstruct $\vec{x}$ exactly, the following certain condition has to be met relating parameters $M$, $N$, and $K$:

$$N > K \log M \qquad (3)$$

This condition is the necessary condition for sparse signal recovery using $l_1$ norm minimization obtained by Donoho and Tanner [6; 7; 8]. For $K \sim 10$ and $M \sim 10^6$ we obtain the following condition for the number of olfactory receptor types necessary for the recovery of the stimulus: $N > 200$, which is satisfied for both humans ($N \approx 350$) and mice ($N \approx 1000$). Thus, equation (1) can in principle be solved with the existing parameters in the olfactory system. This means that the olfactory system can reconstruct $K \sim 10$ monomolecular components given the responses of $N \approx 1000$ olfactory receptors.

The solution (2) is limited in that it is not concentration invariant. Doubling the concentration vector $\vec{x}$ results in the doubling of the vector of receptor responses $\vec{y} = A\vec{x}$. Solving equation (2) under the constraint of doubled vector $\vec{y}$ in turn reconstructs the doubled vector of concentrations $\vec{x}$. Our goal, however, is to build a representation that is concentration invariant. This means that we would like to obtain a framework in which doubling of the receptor responses $\vec{y}$ does not affect the reconstructed stimulus, i.e. a concentration-invariant odorant identity. We therefore propose a primacy-based decoding model that makes this possible.

**2.3 Concentration invariant decoding algorithm via primacy.** To make odorant representation invariant to concentration, we propose to reformulate the sparse recovery problem. In its simplest form, for each input vector $\vec{x}$, and the resulting receptor response vector $\vec{y} = A\vec{x}$, we propose to isolate the set of $p$ strongest responding receptor neurons ($p \ll 1000$). We call this set the primacy set $P$. The complementary set $\bar{P}$ includes weaker responding receptor types. For these two sets we can state that

$$y_P \geq y_{\bar{P}} \qquad (4)$$

Here $y_P$ means the set of components of vector $\vec{y}$ that belong to the primacy group $P$. With these constraints, the sparse recovery problem (2) can be reformulated as follows:

$$\vec{x} = \underset{\substack{y_P \geq y_{\bar{P}} \\ x_j \geq 0}}{\arg\min} \sum_j x_j \qquad (5)$$

When seeking a minimum in this equation, we have to use the relationship $\vec{y} = A\vec{x}$, however, we do not use knowledge about exact values of the vector $\vec{y}$, only the relationships between its components. This approach can be used to recover the concentration vector $\vec{x}$, however, the amplitude of this vector does not depend on the absolute values of the neural responses $y_j$. This is due to the non-parametric formulation of relative rather than absolute values of receptor responses in the sparse signal recovery problem (5). For example, doubling each component of vector $\vec{y}$ does not change the set of inequalities $y_P \geq y_{\bar{P}}$, and, thus, the reconstructed stimulus $\vec{x}$ is not affected. This decoding scheme therefore yields a reconstruction of the stimulus $\vec{x}$ that is concentration invariant: many $\vec{x}$ stimuli with the same direction and different length will yield a reconstructed concentration vector $\vec{x}$ that is exactly the same. We therefore suggest that the primacy constraints result in recovery of concentration-invariant stimuli, leading to the decoding of the concentration-invariant odorant identity.

Similarly, computing a monotonically increasing non-linear function of each component of $\vec{y}$, such as that given by equation (1), does not affect the reconstructed vector $\vec{x}$. Such an operation does not affect the constraint $y_P \geq y_{\bar{P}}$, thus leading to the same minimum $l_1$ norm $\vec{x}$. Within our approach the odorant identity is therefore invariant with respect to the non-linear monotonic transformations of receptor responses. One can therefore use linear and non-linear models of receptor responses i.e. $\vec{y}$ and $\vec{r}$ interchangingly, since this choice does not affect the solution. Overall, using relative rather than absolute values of sensor responses to reconstruct stimuli makes reconstruction invariant to a set of transformations, such as stimulus intensity etc. Although we demonstrated this idea for the particular example of primacy coding, we propose that this *neural relativity rule* could be used more generally by neural systems to transmit and recover signals invariant to various transformations. Instead of minimizing the $l_1$ norm alone to find a sparse solution, one can minimize the elastic net functional:

$$\vec{x} = \arg\min_{\substack{y_P \geq y_{\bar{P}} \\ x_j \geq 0}} \sum_j \left( x_j + \varepsilon \frac{x_j^2}{2} \right) \qquad (6)$$

The equation (6) is the same as the $l_1$ norm minimization (5) when the parameter $\varepsilon \to 0$. For sufficiently small $\varepsilon$, equation (6) yields sparse solutions similar to equation (5). Problem (6), however allows a more straightforward formulation in the dual space. We will therefore use equation (6), i.e. elastic net minimization, to recover olfactory stimuli for the rest of the paper, with a sufficiently small $\varepsilon$.

**2.4 Number of receptors needed to implement primacy coding.** To estimate the number of primary receptors $p$ needed to recover the stimulus we estimate a possible number of combinations in $\vec{y}$ that differ in their primacy code $\Gamma_y \sim C_N^p \sim N^p$. The estimate for the number of combinations in $\vec{x}$ is $\Gamma_x \sim M^K$. Assuming that for an exact reconstruction of $\vec{x}$ the number of combinations of $\vec{y}$ has to exceed the number of combinations in $\vec{x}$, we arrive to the following condition:

$$p > K \log M / \log N \qquad (7)$$

For a typical mammal, such as a mouse, for which the number of olfactory receptors is approximately $N \approx 1000$, we obtain $p > 2K$. For humans, the number of functional olfactory receptors is $N \approx 350$ and, therefore the primacy number (the number of primacy receptor types) required is somewhat higher, $p > 2.4K$. In both of these estimates we assumed that the potential number of types of molecules available in the environment is close to $M \sim 10^6$. If one assumes that the number of discernable molecular components in each mixture is $K \sim 12$, as follows from human psychophysics [5], we obtain the primacy number $p > 30$ which is substantially less than the total number of receptors $N \approx 1000$.

**2.5 Duality transformation.** Since problem (6) includes various inequality constraints, one can use the Lagrange multiplier method to solve it. To present the problem in Lagrange form, one has to reformulate the conditions (4) in a more convenient form. Indeed, minimization of $l_1$ norm or elastic net functional of $\vec{x}$ connected to $\vec{y}$ via a set of positive coefficients $A$ ($\vec{y} = A\vec{x}$) results in the solution $\vec{x} = 0$. To obtain a non-zero solution, one has to introduce a finite scale into the conditions (4). A set of equivalent conditions with constrained magnitude is

$$y_P \geq \gamma, \; \gamma \geq y_{\bar{P}} \qquad (8)$$

where $\gamma \geq 0$ is the scale parameter. By introducing a sign variable $u_i = +1$ for $i \in P$ and $u_i = -1$ for $i \in \bar{P}$, equation (8) can be rewritten as a single set of conditions

$$u_i(y_i - \gamma) \geq 0 \tag{9}$$

Here $y_i = \sum_j A_{ij} x_j$ with receptor $i$ to odorant $j$ affinities given by a set of non-negative numbers $A_{ij}$. Problem (6) combined with constraints (9) represents the primal optimization problem.

To transform the primal minimization problem (6) to the dual problem, we introduce the Lagrangian with two sets of Lagrange multipliers, $\alpha_i$ and $\beta_j$. The former multipliers, $\alpha_i$, with $i = 1..N$ correspond to the constraints (9), while the latter, $\beta_j$, enforce the constraint $x_j \geq 0$ for $j = 1..M$. The full Lagrangian for the elastic net minimization problem (6) is as follows

$$L(\vec{x}, \vec{\alpha}, \vec{\beta}) = \sum_j (x_j + \varepsilon x_j^2/2) - \sum_i \alpha_i u_i (\sum_j A_{ij} x_j - \gamma) - \sum_j \beta_j x_j \tag{10}$$

In this equation, the Lagrange multipliers $\alpha_i$ and $\beta_j$ are supposed to have non-negative values, i.e. $\alpha_i, \beta_j \geq 0$ [9]. To transform the primal problem into the dual problem, we minimize the Lagrangian (10) with respect to $\vec{x}$ if no constraints were present to compute the optimal value of $\vec{x}$ denoted here as $\vec{x}^*$. We then compute the value of the Lagrangian in the minimum $\theta(\vec{\alpha}, \vec{\beta}) \equiv L(\vec{x}^*, \vec{\alpha}, \vec{\beta})$:

$$\partial L(\vec{x}, \vec{\alpha}, \vec{\beta})/\partial \vec{x} = 0 \rightarrow x_j^* = (\sum_i \alpha_i u_i A_{ij} + \beta_j - 1)/\varepsilon \tag{11}$$

$$\theta(\vec{\alpha}, \vec{\beta}) = -\frac{1}{2\varepsilon} \sum_{ik} \alpha_i u_i G_{ik} \alpha_k u_k - \frac{1}{\varepsilon} \sum_{ij} \alpha_i u_i A_{ij} \beta_j - \frac{1}{2\varepsilon} \sum_j (\beta_j^2 - 2\beta_j) + \\ + \sum_i \alpha_i u_i (\gamma + \frac{1}{\varepsilon} \sum_j A_{ij}) \tag{12}$$

Here $G_{ik} = \sum_j A_{ij} A_{kj}$ in the Gramm matrix for rows of matrix $\hat{A}$. According to optimization theory [9] the dual Lagrangian $\theta(\vec{\alpha}, \vec{\beta})$ is to be *maximized* to find the optimal values of $\vec{\alpha}$ and $\vec{\beta}$. These values can be used to find the solution of the primal problem using equation (11). The dual problem can therefore be formulated as follows

$$(\vec{\alpha}^*, \vec{\beta}^*) = \arg\max_{\alpha_i \geq 0, \beta_j \geq 0} \theta(\vec{\alpha}, \vec{\beta}) \tag{13}$$

The reason why the dual problem (13) is preferable to the primal problem (6) lies in the simplicity of the constraints. The constraints $\alpha_i, \beta_j \geq 0$ are easier to implement than inequalities (9). The main observation that we make in this study is that, for the nervous system, it is especially easy to impose non-negativity constraints, because neural responses are described by firing rates that cannot fall below zero. Motivated by this observation, we will argue below that Lagrange multipliers $\vec{\alpha}$ and $\vec{\beta}$ could be represented by the responses of different types of olfactory neurons that solve the dual rather than primal representation problem.

Another motivation for relating $\vec{\alpha}$ and $\vec{\beta}$ to neural activity can be derived from the Karush-Kuhn-Tucker theorem (KKTT) [9]. According to KKTT, at the allowed maximum of the dual Lagrangian (12), the Lagrangian contributions in equation (10) vanish, i.e.

$$\alpha_i u_i (\sum_j A_{ij} x_j - \gamma) = 0, \quad \beta_j x_j = 0 \tag{14}$$

The former equality means that if $\sum_j A_{ij} x_j - \gamma > 0$ (inactive constraint), the corresponding Lagrange multiplier $\alpha_i = 0$. KKTT therefore has significant implications from the standpoint of neural representations. That some values of $\alpha_i$ are identically equal zero (for inactive constraints) makes the vector of responses sparse, which is one of the broadly observed features of neural responses [10; 11; 12] including in the olfactory system [13; 14; 15; 16]. The mapping between the dual problem and neural responses, if established, could explain the sparsity of neural responses as a corollary of KKTT.

**2.6 Dual network.** We will now describe the neural networks that solve the dual problem (12) and (13). We associate vectors $\vec{\alpha}$ and $\vec{\beta}$ with firing rates of two groups of neurons (cell types). The conditions $\alpha_i \geq 0$ and $\beta_j \geq 0$ are then satisfied automatically as firing rates cannot be negative. We then assume that the neurons are connected into a network and that the Lyapunov function of the network $H(\vec{\alpha}, \vec{\beta})$ is proportional to the negative dual Lagrangian: $H(\vec{\alpha}, \vec{\beta}) = -\varepsilon \theta(\vec{\alpha}, \vec{\beta})$ [cf. equation (12)]. To generate network equations from the Lyapunov function, for each neuron in the network, we define internal variables that can be viewed as total synaptic input current, for example. For $\alpha_i$ and $\beta_j$ units, these currents will be denoted $a_i$ and $b_j$ respectively. Consider the following equations for $a_i$ and $b_j$

$$\dot{a}_i + a_i = -\sum_k W_{ik} \alpha_k - \sum_j u_i A_{ij} \beta_j + \varepsilon \gamma u_i + u_i \sum_j A_{ij} \tag{15}$$

$$\dot{b}_j + b_j = -\sum_i \alpha_i u_i A_{ij} + 1 \tag{16}$$

$$W_{ik} = u_i G_{ik} u_k - \delta_{ik} \tag{17}$$

$$\alpha_i = [a_i]_+, \quad \beta_j = [b_j]_+ \tag{18}$$

Here and throughout this paper we use $\dot{x} = dx/dt$. The first equation describes inputs into cells with firing rates $\alpha_i$ connected by weights $-W_{ik}$. Connectivity between $\alpha$ cells is symmetric. These cells are also connected to $\beta_j$ cells with synaptic weights $-u_i A_{ij}$. $\alpha$ cells also receive external excitatory drive equal to $\varepsilon \gamma u_i + u_i \sum_j A_{ij}$. Equation (16) describes $\beta_j$ cells that are connected symmetrically to $\alpha$ cells. For both cell types, their firing rates ($\alpha_i$ and $\beta_j$) are connected to their inputs ($a_i$ and $b_j$) by rectifying threshold-linear relationships (18) ($[x]_+ = x$ for $x \geq 0$ and $[x]_+ = 0$ for $x < 0$). It is straightforward to show that these equations can be rewritten as forms of gradient descend

$$\dot{a}_i = -\partial H(\vec{\alpha}, \vec{\beta}) / \partial \alpha_i, \quad \dot{b}_j = -\partial H(\vec{\alpha}, \vec{\beta}) / \partial \beta_j \tag{19}$$

To show that $H(\vec{\alpha}, \vec{\beta})$ is indeed the Lyapunov function of equations (15)-(18), i.e. a function that is not increasing when our network evolves according to these equations, we observe that

$$dH(\vec{\alpha}, \vec{\beta})/dt = \sum_i \dot{\alpha}_i \partial H(\vec{\alpha}, \vec{\beta})/\partial \alpha_i + \sum_j \dot{\beta}_j \partial H(\vec{\alpha}, \vec{\beta})/\partial \beta_j =$$
$$= -\sum_i \dot{a}_i \dot{\alpha}_i - \sum_j \dot{b}_j \dot{\beta}_j = -\sum_i f'(a_i) \dot{a}_i^2 - \sum_j f'(b_j) \dot{b}_j^2 \leq 0 \tag{20}$$

Here $f(x) = [x]_+$, $f'(x) \geq 0$. Because $dH(\vec{\alpha}, \vec{\beta})/dt \leq 0$, our network will minimize the Lyapunov function, and, by extension, maximize the dual Lagrangian (12). Because of the constraints imposed on the firing rates (18), these variables will stay non-negative in the course of this optimization, thus automatically satisfying the constraints imposed on the variables of the dual Lagrangian. We conclude therefore that our network, which includes only two cell types, can solve the dual constraint optimization problem, leading to the accurate computation of the molecular composition of a mixture in dual space.

The purpose of variables $u_i$ is to identify the set of primary variables $y_i$, i.e. the set of $p$ components of vector $\vec{y}$ that are larger than all others. To compute these variables one could use the network with winner-takes-all architecture, except one has to ensure that there are $p$ winners. This network can be implemented by connecting $N$ neurons with an unstructured global inhibition, in the manner described previously by us [17]: $\dot{v}_i + v_i = -c\sum_k u_k + y_i$,

$u_i = \text{sgn}(v_i)$. Here, $c$ is the strength of global inhibition. By adjusting parameter $c$, one can ensure that a given number of $u$-cells is active.

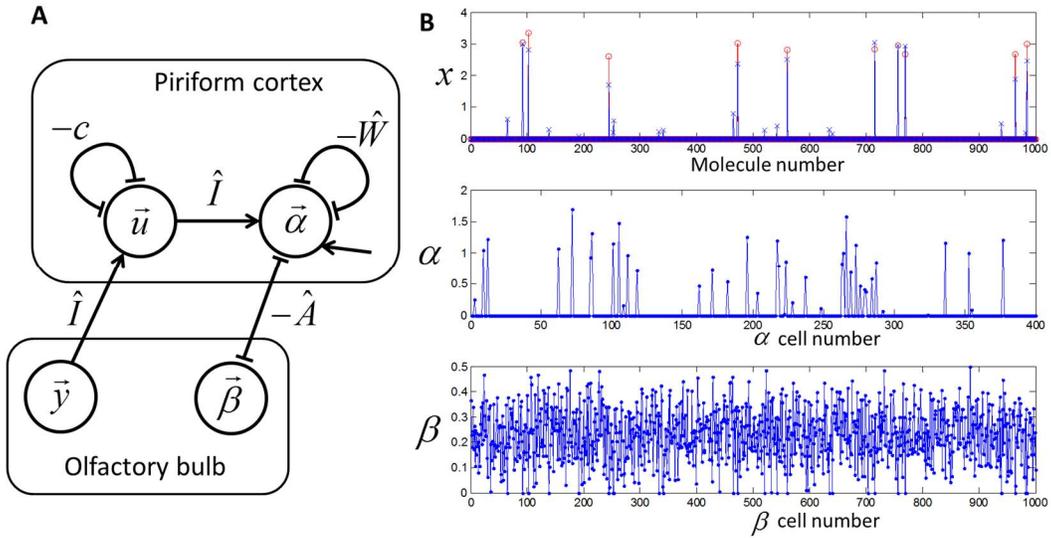

Figure 1. Simulations with the network described. (A) The structure of the network. We propose that $\alpha$ cells implementing the dual representation of the concentration vector reside in the piriform cortex (PC). This is because PC is known to have extensive recurrent connectivity. $\beta$ cells, which implement the non-negativity constraints on the concentration vector, could be related to the granule cells of the olfactory bulb [14; 18]. (B) Firing rate simulation of the dual model (blue) successfully identifies the non-zero components of the stimulus (top, red circles). Small false positive components of $x$ result from our use of the elastic net measure and disappear as $\varepsilon \to 0$. They can be removed by thresholding the reconstruction. In the simulation, we used $p = 60$.

## 3 Discussion

Herein we propose a novel model for the intensity-invariant encoding of olfactory stimuli. According to the primacy model, an odor can be identified on the basis of the identities of the $p$ strongest responding olfactory receptors. In the case of the human olfactory system, we have estimated $p$ to be ~30. Because the primacy model relies on the relative rather than absolute strengths of receptor responses, the recovered stimulus is independent wrt the absolute stimulus concentration and depends only on the relative concentrations of individual molecules in the mixture. Although we demonstrated this idea for the particular example of primacy coding, we suggest that this rule of neural relativity could be used to produce intensity invariant signal recovery for more complex conditions and in other modalities.

We then attempted to formulate the solution of the decoding problem from the first principles. To implement primacy decoding, we cast this problem as a dual Lagrange problem, which required introducing two sets of Lagrange multipliers. The first set ($\alpha$) enforced the primacy conditions, while the second set ($\beta$) ensured the non-negativity of the individual molecular concentrations. By assuming that the dual Lagrangian corresponds to a Lyapunov function of a network, we were able to derive the network structure. Interestingly, we found the $\alpha$ cells were found to be recurrently connected, while $\beta$ cells are connected to $\alpha$ cells via inhibitory circuitry. These features are reminiscent of the cells in real olfactory networks. Pyramidal cells in the piriform cortex (PC) form extensive recurrent connections [19], while granule cells of the olfactory bulb receive inhibitory connections from PC [18]. The connection from the granule cells to the PC is indirectly inhibitory, because granule cells inhibit the mitral cells within the olfactory bulb through dendrodendritic synapses. Mitral cells are excitatory cells projecting to PC. These considerations allow us to place $\alpha$ cells and related $u$ cells into the piriform cortex and associate $\beta$ cells with the granule cells of the olfactory bulb. Interestingly, the number of $\beta$ cells matches the number of monomolecular components that the system can resolve, because these cells implement the conditions of non-negativity of the molecular concentrations. The number of granule cells found in the olfactory bulb is a few million, which matches with our estimate of the number of volatile molecules [18]. This coincidence provides a further argument in favor of the association of $\beta$ cells with granule cells.

Lengyel et al. [20] have proposed that the maximum *a posteriori* (MAP) solution to the inference problem of recovering a sparse N-dimensional odor vector x can be achieved in a low-dimensional measurement space, reflecting the known biology of olfactory processing. In their study, the compressed sensing problem was formulated as an $l_1$ norm minimization of the odor vector x subject to the linear equality encoding constraint y = Ax and solved by considering the problem in dual space. The resulting generalized energy function for the network reflects the equality constraints and, as such, their proposed network implementation differs significantly from our solution, which relies on Karush-Kuhn-Tucker-type inequality conditions implementing primacy.

Two features of dual networks are worth highlighting. First, according to Karush-Kuhn-Tucker theory, dual Lagrangians are optimized under constraints of non-negativity of the Lagrange coefficients. Neuronal responses can naturally enforce these constraints, because they are described by firing rates that cannot fall below zero. Secondly, due to KKTT [9], a large number of the Lagrange coefficients are bound to be zero, drawing an interesting parallel with SVMs. This observation is consistent with the observed sparsity of neuronal responses, both in olfaction [13; 14; 15; 16] and beyond [10; 11; 12; 21], further strengthening the possible association between neuronal responses and Lagrange coefficients. We therefore provide arguments in favor of relating neuronal responses and Lagrange multipliers implementing several constraints present in the stimulus.